\documentclass[twocolumn]{aastex6}
\usepackage{amsmath}
\usepackage{amssymb}
\usepackage{latexsym}
\usepackage{listings}
\usepackage{graphicx}
\usepackage{rotating}
\usepackage{url}
\usepackage{theorem}
\usepackage{verbatim}
\usepackage{natbib}
\DeclareGraphicsExtensions{.pdf}
\shorttitle{Dusty Stars in Moving Groups from Disk Detective}
\shortauthors{Silverberg et al.}

\begin{document}

\title{A New M Dwarf Debris Disk Candidate in a Young Moving Group Discovered with Disk Detective}
\author{Steven M. Silverberg\altaffilmark{1}, 
Marc J. Kuchner\altaffilmark{2}, 
John P. Wisniewski\altaffilmark{1}, 
Jonathan Gagn\'e\altaffilmark{3,4},
Alissa S. Bans\altaffilmark{5},
Shambo Bhattacharjee\altaffilmark{6}, 
Thayne R. Currie\altaffilmark{7},
John R. Debes\altaffilmark{8},
Joseph R. Biggs\altaffilmark{9}, 
Milton Bosch\altaffilmark{9}, 
Katharina Doll\altaffilmark{9}, 
Hugo A. Durantini-Luca\altaffilmark{9}, 
Alexandru Enachioaie\altaffilmark{9},
Philip Griffith, Sr.\altaffilmark{9},
Michiharu Hyogo\altaffilmark{9},
Fernanda Pi\~niero\altaffilmark{9},
Disk Detective Collaboration\altaffilmark{10}
}
\email{silverberg@ou.edu}
\altaffiltext{1}{Homer L. Dodge Department of Physics and Astronomy, University of Oklahoma, 440 W. Brooks St., Norman, OK 73019, USA}
\altaffiltext{2}{NASA Goddard Space Flight Center, Code 667, Greenbelt, MD 20771, USA}
\altaffiltext{3}{Department of Terrestrial Magnetism, Carnegie Institution of Washington, 5241 Broad Branch Road NW, Washington, DC 20015-1305, USA}
\altaffiltext{4}{NASA Sagan Fellow}
\altaffiltext{5}{Adler Planetarium, 1300 S Lake Shore Dr., Chicago, IL 60605}
\altaffiltext{6}{School of Statistics, University of Leeds, Leeds LS2 9JT, England}
\altaffiltext{7}{National Astronomical Observatory of Japan, Subaru Telescope}
\altaffiltext{8}{ESA/AURA Astronomer, Space Telescope Science Institute, 3700 San Martin Drive, Baltimore, MD 21218, USA}
\altaffiltext{9}{Disk Detective}
\altaffiltext{10}{\url{http://www.diskdetective.org/\#/authors}}

\begin{abstract}
We used the Disk Detective citizen science project and the BANYAN II Bayesian analysis tool to identify a new candidate member of a nearby young association with infrared excess.  WISE J080822.18-644357.3, an M5.5-type debris disk system with significant excess at both 12 and 22 $\mu$m, is a likely member ($\sim 90\%$ BANYAN II probability) of the $\sim 45$ Myr-old Carina association. Since this would be the oldest M dwarf debris disk detected in a moving group, this discovery could be an important constraint on our understanding of M dwarf debris disk evolution.
%
%
\end{abstract}

 
\section{Introduction}
Young moving groups and associations (YMGs) provide highly valuable targets for exoplanet searches and important tracers of disk evolution \citep{ZuckermanSong2004, LopezSantiago2006}. The stars in YMGs share similar ages, and their motions through the Galaxy trace back to a common origin locus. Exoplanets around stars in YMGs are often young and warm enough to be observed via near-infrared direct imaging with large telescopes \citep[e.g.][]{Marois2008, Currie2014}. Additionally, the well-determined ages of YMG members allow us to place any planets and disks discovered around them in a chronological sequence, tracing the evolution of planetary systems.




YMGs have driven many studies of disk evolution. Based on \textit{Spitzer} data and disk models of objects in the Taurus cloud and Ophiuchus, \citet{Espaillat2010} proposed that the gaps in pre-transitional disks were indicators of planet formation. Mid- (24 $\mu$m) and far-IR (70 $\mu$m) observations of moving group A stars from \textit{Spitzer}/MIPS show that disks around older stars have a narrower temperature distribution than younger stars, and that 70 $\mu$m disk emission persists longer than 24 $\mu$m emission \citep{Su2006}. \citet{KennedyWyatt2010} modeled the potential self-stirring behavior of disks and applied it to A-stars in the $\beta$ Pictoris moving group and the TW Hya association.  Further observations and analysis of archival data of the Tucana/Horologium, AB Doradus, Columba, and Argus associations with \textit{Spitzer} \citep{Zuckerman2011} characterized the appearance of debris disks around YMG stars, and constrained the decay of dusty debris disks over time.


In recent years, many direct-imaging exoplanet and disk surveys have focused on YMG members. \citet{Kasper2007} primarily surveyed the Tucana-Horologium and $\beta$ Pictoris moving groups in their direct imaging search for Jupiter-mass sub-stellar companions.  \citet{Biller2013} directly imaged 80 members of the $\beta$ Pictoris, TW Hya, Tucana-Horologium, AB Doradus, and Hercules-Lyrae YMGs in a survey for giant planets as part of the Gemini/NICI Planet-Finding Campaign, finding four co-moving companions and constraining the frequency of 1-20 $M_{\mathrm{Jup}}$ planets at separations up to 150\,AU. The Strategic Exploration for Exoplanets and Disks with Subaru (SEEDS) survey included a dedicated focus on potential moving group members \citep{Brandt2014}, using high-contrast coronagraphic imaging to observe many targets identified with YMGs. \citet{Currie2015} identified a Kuiper belt-like debris disk in the Scorpius-Centaurus association with the Gemini Planet Imager as part of a larger survey.




The Wide-field Infrared Survey Explorer (WISE) mission \citep{Wright2010}, the most sensitive all-sky mid-infrared survey to date, provides the best source of new YMG candidate members with infrared excess across the sky. Some past WISE studies to identify late-type YMG candidate members with infrared excess have generally proceeded by first identifying likely members, then examining WISE data to see if they have infrared excesses \citep[e.g.][]{Schneider2012a, Schneider2012b}. Other WISE studies of YMG candidate membership of earlier-type stars have used the presence and strength of infrared excess in WISE data as a component in their membership determination \citep[e.g.][]{Rizzuto2012}. We instead first identify stars with infrared excesses, and then test each one for membership in a YMG.

We report here one newly identified candidate YMG member with infrared excess discovered by the Disk Detective project \citep[][hereafter Paper 1]{Kuchner2016}. This star has $>90\%$ membership probability in a YMG based on the BANYAN II Bayesian analysis tool \citep{Malo2013, Gagne2014}. Distance estimates ($<88$ pc) suggest that this star is a good target for direct imaging. We also recover a previously-identified YMG candidate member with a known infrared excess.

In Section 2 of this paper, we summarize our methodology for identifying and modeling stars with excesses and determining their YMG membership probability in more detail. In Section 3, we present our candidate member, discuss its stellar characteristics, compare its kinematics to its association, and characterize basic parameters of the star and its infrared excess. In Section 4, we summarize our findings and discuss the outlook for future identifications in this manner, in light of the upcoming release of results from the \textit{Gaia} mission \citep{Perryman2001}.

\section{Methodology}

The objects with identified infrared excesses come from the Disk Detective project (Paper 1; \url{http://www.diskdetective.org}), a citizen science-based all-sky search for circumstellar disks in the AllWISE Data Release \citep{Cutri2014}. Unlike other \textit{WISE} debris disk searches, Disk Detective is not limited to the \textit{Hipparcos} or \textit{Tycho} catalogs, which are magnitude-limited in $V$ and thus omit a wide swath of mid- and late-type stars. It instead searches the entire 2MASS catalog \citep{Skrutskie2006} for objects with [W1] - [W4] $> 0.25$ that meet other criteria designed to eliminate contaminants and other sources of noise. The DiskDetective.org website aims to harvest Disk Detective Objects of interest (DDOIs), sources we consider to be worthy of further research. DDOIs are then submitted for follow-up observation on ground- and space-based telescopes. Details on the identification of DDOIs can be found in Paper 1. As of 25 April, 2016, DiskDetective.org users had identified 1774 DDOIs, a unique new collection of potential debris disks.

To test the likelihood of YMG membership for these objects, we used Bayesian Analysis for Nearby Young AssociatioNs II \citep[BANYAN II;][]{Malo2013, Gagne2014}. This tool uses a naive Bayesian classifier to compare the Galactic position and space velocity of a given object to the positions and velocities of several well-defined moving groups and associations with distances $<100$ pc and ages $<200$ Myr: the $\beta$ Pictoris, AB Doradus, and TW Hya moving groups, as well as the Argus, Columba, Carina, and Tucana-Horologium associations. BANYAN II takes as inputs an object's right ascension, declination, trigonometric parallax, radial velocity, and proper motion. We tested 1774 DDOIs with BANYAN II, using spatial coordinates from AllWISE, and parallax and proper motions from \textit{Hipparcos} \citep{Hipparcos} or the Tycho-Gaia Astrometric Solution \citep[TGAS,][]{Lindegren2016} whenever available. When  parallax and proper motion data from these surveys were not available, we used proper motion data from other surveys, such as the Tycho-2 bright star catalog \citep{Hog2000}, or the SPM4 catalog \citep{Girard2011}.

We used an algorithm based on the implementation of the Levenberg-Marquardt minimization scheme in the Python \texttt{lmfit} package to estimate basic parameters of the host star and disk of our YMG candidates. Our algorithm cycles through combinations of $T_{\mathrm{eff}}$ and $\log(g)$, fitting the corresponding BT-Settl stellar atmosphere model \citep{Baraffe2015} to 2MASS $J$, $H$, and $K$, WISE W1, and any additional large-survey photometry (e.g. the Tycho2 catalog or the DENIS survey \citep{Epchtein1997}) to find a best-fit ratio of stellar radius to distance. We adopt the best-fitting of these models, yielding estimates of the stellar temperature $T_{\star}$ and distance. Taking this distance and stellar model into account, the algorithm then fits the four bands of WISE photometry with a Planck function, representing a single-temperature disk. The parameters from these two fits are then used to derive a value for the disk's fractional infrared luminosity, $L_{ir}/L_{\star}$.

\begin{deluxetable*}{lccccccc}
\tablecaption{Summary of a Candidate YMG Member from Disk Detective
\label{table:results}}
\tablehead{ & \colhead{Spectral} & & \colhead{Moving} & \colhead{Membership} & \colhead{Group} & \colhead{Most Probable\tablenotemark{2}} & \colhead{Most Probable\tablenotemark{2}}\\
\colhead{ID} & \colhead{Type} & \colhead{$L_{ir}/L_{\star}$} & \colhead{Group} & \colhead{Probability (\%)} & \colhead{Age (Myr)\tablenotemark{1}} & \colhead{Distance (pc)} & \colhead{Radial Velocity ($\mathrm{km\, s^{-1}}$)}}
\startdata
J080822.18-644357.3 & M5.5\tablenotemark{3} & $8.06 \times 10^{-2}$ & Carina & 93.9 & $45^{+11}_{-7}$ & $65.4^{+8.8}_{-7.6}$ & $20.6 \pm 1.4$\\
\enddata
\tablenotetext{1}{From \citet{Bell2015}}
\tablenotetext{2}{From BANYAN II}
\tablenotetext{3}{This work}
\end{deluxetable*}

\section{Results}

From our initial sample of 1774 DDOIs, we identified one star with $>90\%$ likelihood of membership in a YMG, with significant excess [W1]-[W4]$>0.25$. As this system was not included in either the \textit{Hipparcos} survey, or the late-M surveys for BANYAN \citep{Malo2013, Gagne2014}, its membership candidacy has not been evaluated until now. We initially identified WISE J060652.79-313054.1, an F8 star, as a member of the Columba association, based on position and proper motion measurements. and estimated the distance to the star via spectroscopic parallax, independent of the BANYAN II calculation. However, when we incorporated the parallax measurement from TGAS into the BANYAN II analysis, the membership probability dropped from 93.77\% to 0.73\%, indicating that the detection was due to incomplete information. We also recovered TWA 33, which was previously reported in \citet{Schneider2012b}. Table \ref{table:results} summarizes our findings. Here, we briefly discuss the characteristics of this new dusty YMG member.

\subsection{J080822.18-644357.3}

\begin{figure*}[htb]
\begin{centering}
\plottwo{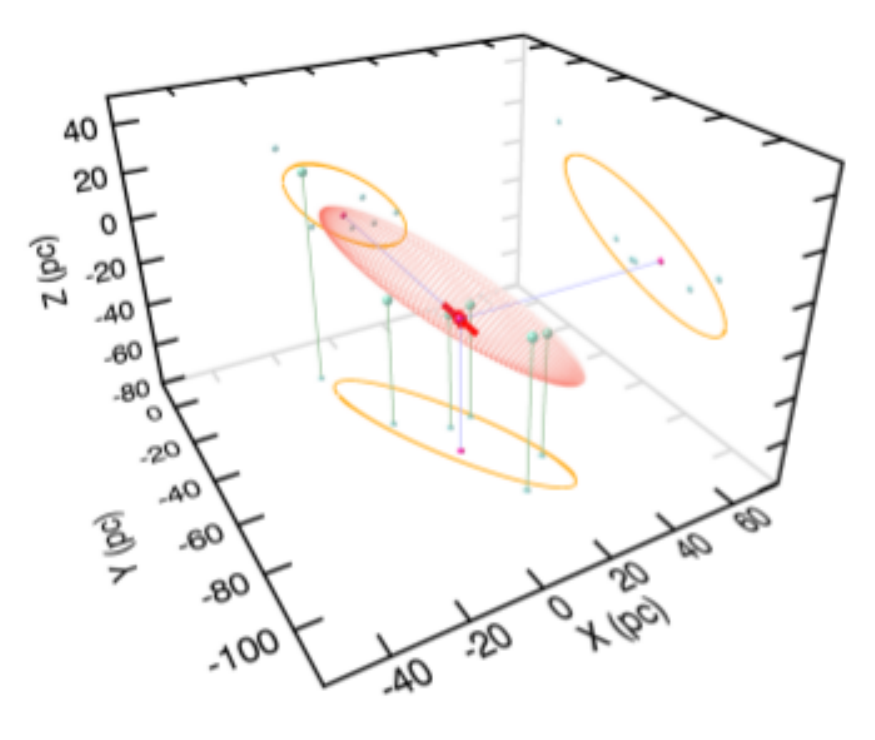}{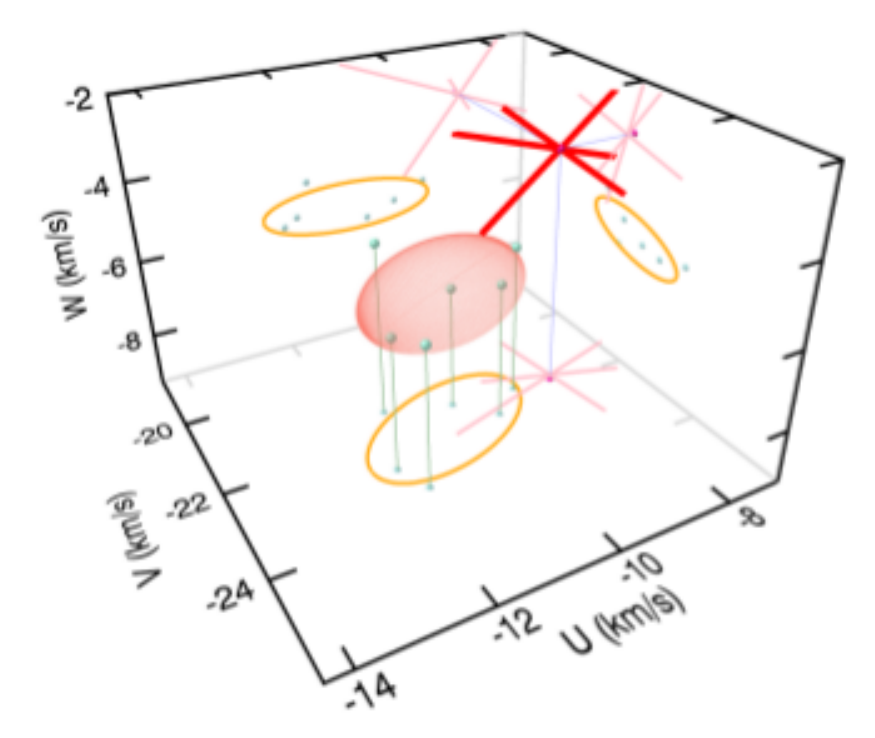}
\caption{Galactic space position and velocity coordinates for J080822.18-644357.3 (purple dot), relative to the members of the Carina association used in BANYAN II (green dots). Red lines are 1$\sigma$ error bars, oriented to decouple errors in proper motion and radial velocity. The orange ellipsoid highlights the 68\% confidence interval in 3D space, while the gold contours indicate the two-dimensional projections of the ellipsoid in each plane. The Galactic position and velocity coordinates for J080822.18-544357.3 indicate that it has  $93.9\%$ probability of membership in this association.}
\label{fig:Star3posvel}
\end{centering}
\end{figure*}

This star has no previously-reported parallax or radial velocity measurements. It does have proper motion observations from the Southern Proper Motion 4.0 (SPM4.0) survey \citep{Girard2011}. Based on this data, it has a 93.9\% probability of membership in the Carina association. This association was first identified as part of the Great Austral Young Association by \citet{Torres2003, Torres2006}, but was later identified as a separate association by \citet{Torres2008}. The association has age $45^{+11}_{-7}$ Myr \citep{Bell2015}, and distance range 46-88 pc \citep[][and references therein]{Malo2013}. Examination of the star's Galactic position and velocity relative to that group (shown in Figure \ref{fig:Star3posvel}) indicates that while it is centrally located in the group at the most probable radial velocity and parallax from BANYAN II, its velocity is not as typical; its three-dimensional velocity falls outside the $68\%$ confidence bubble for the group, indicated by the faint red ellipsoid in Figure \ref{fig:Star3posvel}. However, it is not a particularly extreme outlier; the $1\sigma$ error bars on the star's velocity intersect with the ellipsoid. 

\begin{figure}[tb]
\begin{centering}
\includegraphics[width=0.5\textwidth]{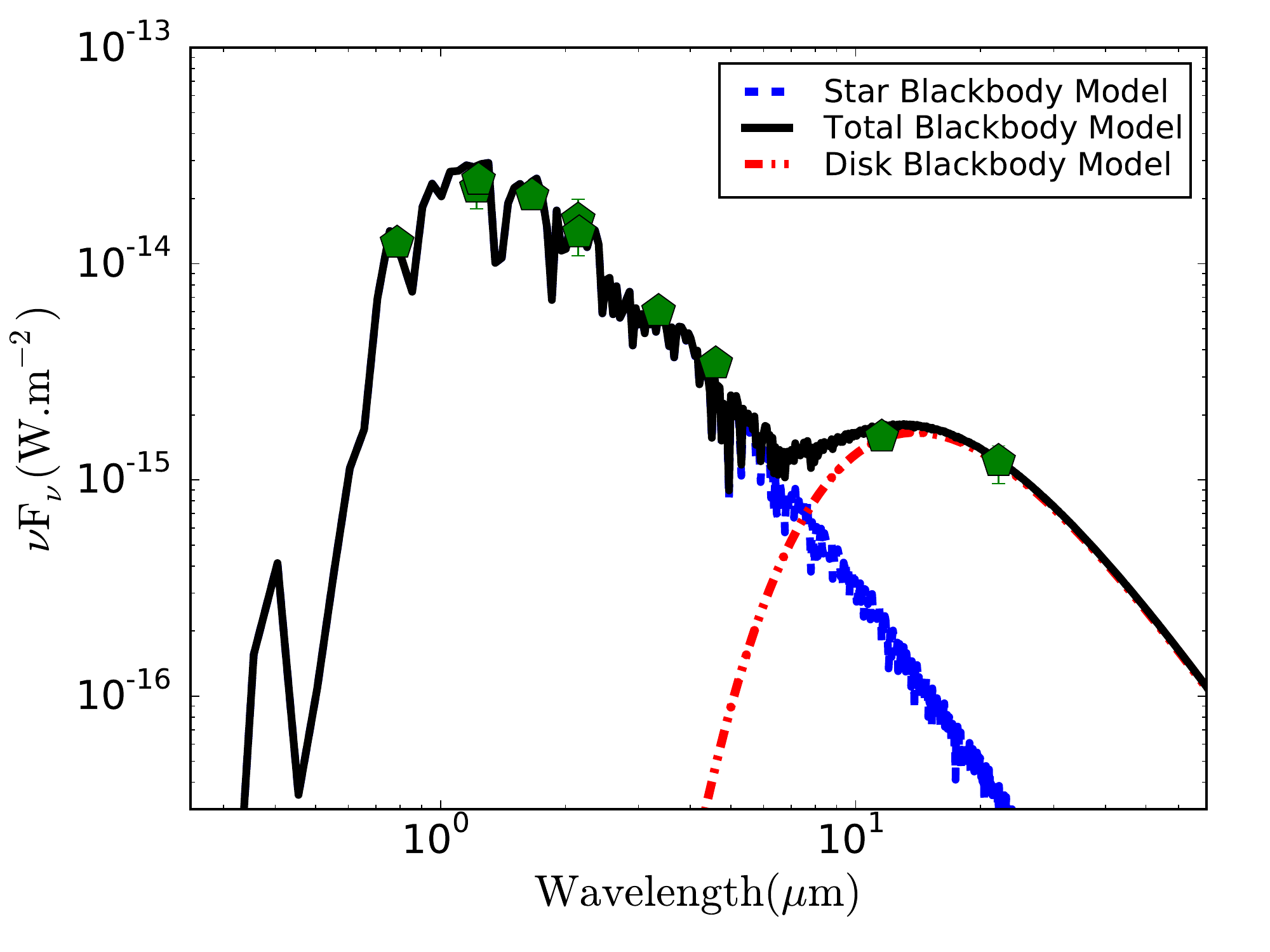}
\caption{Spectral energy distribution for J080822.18-644357.3, with stellar atmosphere (blue dashed line) + blackbody (red dot-dash line) fitting applied to observed photometry (green pentagons) from 2MASS, AllWISE, and the DENIS survey, to produce a total model (black solid line). Our fitting indicates a disk temperature $\sim 263$ K, and $L_{ir}/L_{\star} = 8.06 \times 10^{-2} \pm 9.02 \times 10^{-3}$.}
\label{fig:Star3SED}
\end{centering}
\end{figure}

As seen in Figure \ref{fig:Star3SED}, this star exhibits a significant excess at both W3 ([W1] - [W3] $= 2.367 \pm 0.036$) and W4 ([W1] - [W4] $= 4.234 \pm 0.087$), suggesting a fairly warm disk around a fairly cool star. While this star does not have a previously-reported spectral type, we use photometry from 2MASS, AllWISE, and the DENIS survey \citep{Epchtein1997} to find a best-fit BT-Settl stellar atmosphere with $T_{\star} = 2900$ K, consistent with a spectral class of M5.5V based on the models used by \citet{Rajpurohit2013}. Using this adopted spectral type and an absolute 2MASS $J$ magnitude for this model of 9.21, we find a distance based on spectroscopic parallax of $\sim 57$ pc, which reasonably agrees with the most probable distance from BANYAN II. We find values for the disk of $T_{\mathrm{disk}} \sim 263$ K, and $L_{ir}/L_{\star} = 8.06 \times 10^{-2} \pm 9.02 \times 10^{-3}$. These high values of $L_{ir}/L_{\star}$ and $T_{\mathrm{disk}}$ suggest a young star, independent of its YMG candidacy. The high blackbody temperature of the disk also suggests a close-in disk, with an inner disk  radius of $\sim 0.074$ AU, approximately $1.5$ times the semi-major axis of the orbit of Proxima Centauri b \citep{AngladaEscude2016}.

The excess observed around this star is much larger than the warm debris excess observed around any other mid-M dwarf. The infrared colors of the star are quite similar to the colors of T Tauri stars observed in the youngest embedded clusters \citep[e.g.][]{LuhmanMamajek2012}. Additionally, our derived value of $L_{ir}/L_{\star}$ for this star is comparable to those observed for protoplanetary disks in Taurus, IC 348, and other new star-forming regions. However, large surveys of disks in young associations suggest that gas-rich protoplanetary disks dissipate by $\sim 10$ Myr \citep[e.g.][]{WilliamsCieza2011}. With such a large excess at an age of $\sim 45$ Myr, this disk system is an outlier, which compels further study.



\section{Summary and Discussion}
In this paper, we identified one new star with infrared excesses that has a high ($>90\%$) likelihood of membership in the Carina association. This new YMG disk candidate is a valuable target for further follow-up observations with adaptive-optics systems on large telescopes. As it is a candidate member of a nearby (distances $< 100$ pc) association \citep[and references therein]{Malo2013}, it is well within range for high-contrast imaging to resolve the disk structure \citep{Currie2015,Boccaletti2015,Schneider2014}, which will contribute to our understanding of disk evolution at the age of the Carina association. Its known young age also makes it a prime target for finding exoplanets via direct imaging, given the expected warmth of any exoplanets orbiting the disk \citep{Marois2008,Marois2010}.



If confirmed as a member of Carina, our new debris disk appears to be the oldest observed YMG M dwarf debris disk. The frequency of M dwarf debris disks at varying ages is a subject of intense debate in the literature. The frequency of debris disks around young ($\leq 40$ Myr) M dwarfs is $\sim 6\%$ \citep{Binks2016}, while the prevalence around older M dwarfs is $\leq 1.3\%$ \citep{Avenhaus2012, TheissenWest2014}. In contrast, debris disks are detected around $32 \pm 5\%$ of young A stars with \textit{Spitzer}/MIPS \citep{Su2006}, and around $1-6\%$ of old ($\sim 670$ Myr) Sun-like (F5-K9) stars with \textit{Spitzer}/MIPS \citep{Urban2012}. Survival models predict that M dwarf debris disks occur at a similar frequency as disks around Sun-like stars, and that the dearth of detections to date is either due to systems having blackbody-like dust close to their central star, or due to systems having a smaller amount of dust distributed over a larger orbital separation \citep{HengMalik2013}. Alternatively, disk dissipation could be accelerated around these stars due to stellar wind drag \citep{Plavchan2005, Plavchan2009}. Our new M dwarf debris disk would bridge the gap between YMG and field M dwarf disks. Given their common spectral type (both M5.5V), this system could be a young analog for the Proxima Centauri system \citep{AngladaEscude2016}, as well.

We identified this new candidate YMG member with infrared excess via the ongoing Disk Detective project, out of an initial sample of 1774 DDOIs. We expect to find $\sim 12000$ DDOIs by the end of the Disk Detective project, so we expect to find more YMG candidate disks as the project continues. We detected fewer YMG candidate members from this sample than might have been expected, given the results of \citet{KennedyWyatt2013}. We believe that this low detection rate is in part due to Disk Detective's sizable inclusion of objects with distances $> 100$ pc, more distant than any of the moving groups included in BANYAN II. Additionally, some nearby high-proper-motion targets may have originally been classified as ``shifting'' targets rather than good candidates. We have begun re-evaluation of these shifting targets to identify these false negatives. Forthcoming improvements to the BANYAN software and results from the \textit{Gaia} mission \citep{Perryman2001} should lead to substantially higher yields in the future, as well. We plan to add an additional six known YMGs, which are not currently included in any probabilistic membership tool, to the BANYAN software. The data from \textit{Gaia} are likely to yield many previously-undiscovered YMGs, as well as find previously-unidentified members of the currently known moving groups and associations. We expect that \textit{Gaia} will determine parallaxes, proper motions, and radial velocities for $\sim 70\%$ of our current list of DDOIs. Extrapolating to the anticipated 12,000 DDOIs and assuming the use of \textit{Gaia} data, we estimate that we will identify an additional $\sim 15$ candidate members of the moving groups and associations studied here with significant $22 \mu$m excess by the end of the Disk Detective project.

\acknowledgments
We thank the referee for providing comments that helped to improve the content and clarity of this paper. We acknowledge support from grant 14-ADAP14-0161 from the NASA Astrophysics Data Analysis Program. M.J.K. acknowledges funding from the NASA Astrobiology Program via the Goddard Center for Astrobiology.

Development of the Disk Detective site was supported by a grant from the Alfred P. Sloan Foundation. The Zooniverse platform is supported by a Google Global Impact award.

WISE is a joint project of the University of California, Los Angeles, and the Jet Propulsion Laboratory (JPL)/California Institute of Technology (Caltech), funded by NASA. 2MASS is a joint project of the University of Massachusetts and the Infrared Processing and Analysis Center (IPAC) at Caltech, funded by NASA and the NSF. This paper uses data products produced by the OIR Telescope Data Center, supported by the Smithsonian Astrophysical Observatory.

The Digitized Sky Survey was produced at the Space Telescope Science Institute under U.S. Government grant NAG W-2166. The images of these surveys are based on photographic data obtained using the Oschin Schmidt Telescope on Palomar Mountain and the UK Schmidt Telescope. The plates were processed into the present compressed digital form with the permission of these institutions.

This work has made use of data from the European Space Agency (ESA) mission {\it Gaia} (\url{http://www.cosmos.esa.int/gaia}), processed by the {\it Gaia} Data Processing and Analysis Consortium (DPAC, \url{http://www.cosmos.esa.int/web/gaia/dpac/consortium}). Funding for the DPAC has been provided by national institutions, in particular the institutions participating in the {\it Gaia} Multilateral Agreement.


\begin{thebibliography}{}
\providecommand\natexlab[1]{#1}
\providecommand\JournalTitle[1]{#1}

\bibitem[{{Anglada-Escud{\'e}} {et~al.}(2016){Anglada-Escud{\'e}}, {Amado},
  {Barnes}, {Berdi{\~n}as}, {Butler}, {Coleman}, {de La Cueva}, {Dreizler},
  {Endl}, {Giesers}, {Jeffers}, {Jenkins}, {Jones}, {Kiraga}, {K{\"u}rster},
  {L{\'o}pez-Gonz{\'a}lez}, {Marvin}, {Morales}, {Morin}, {Nelson}, {Ortiz},
  {Ofir}, {Paardekooper}, {Reiners}, {Rodr{\'{\i}}guez},
  {Rodr{\#943}guez-L{\'o}pez}, {Sarmiento}, {Strachan}, {Tsapras}, {Tuomi}, \&
  {Zechmeister}}]{AngladaEscude2016}
{Anglada-Escud{\'e}}, G., {Amado}, P.~J., {Barnes}, J., {et~al.} 2016,
  \href{http://dx.doi.org/10.1038/nature19106}{\JournalTitle{\nat}, 536, 437}

\bibitem[{{Avenhaus} {et~al.}(2012){Avenhaus}, {Schmid}, \&
  {Meyer}}]{Avenhaus2012}
{Avenhaus}, H., {Schmid}, H.~M., \& {Meyer}, M.~R. 2012,
  \href{http://dx.doi.org/10.1051/0004-6361/201219783}{\JournalTitle{\aap},
  548, A105}

\bibitem[{{Baraffe} {et~al.}(2015){Baraffe}, {Homeier}, {Allard}, \&
  {Chabrier}}]{Baraffe2015}
{Baraffe}, I., {Homeier}, D., {Allard}, F., \& {Chabrier}, G. 2015,
  \href{http://dx.doi.org/10.1051/0004-6361/201425481}{\JournalTitle{\aap},
  577, A42}

\bibitem[{{Bell} {et~al.}(2015){Bell}, {Mamajek}, \& {Naylor}}]{Bell2015}
{Bell}, C.~P.~M., {Mamajek}, E.~E., \& {Naylor}, T. 2015,
  \href{http://dx.doi.org/10.1093/mnras/stv1981}{\JournalTitle{\mnras}, 454,
  593}

\bibitem[{{Biller} {et~al.}(2013){Biller}, {Liu}, {Wahhaj}, {Nielsen},
  {Hayward}, {Males}, {Skemer}, {Close}, {Chun}, {Ftaclas}, {Clarke}, {Thatte},
  {Shkolnik}, {Reid}, {Hartung}, {Boss}, {Lin}, {Alencar}, {de Gouveia Dal
  Pino}, {Gregorio-Hetem}, \& {Toomey}}]{Biller2013}
{Biller}, B.~A., {Liu}, M.~C., {Wahhaj}, Z., {et~al.} 2013,
  \href{http://dx.doi.org/10.1088/0004-637X/777/2/160}{\JournalTitle{\apj},
  777, 160}

\bibitem[{{Binks}(2016)}]{Binks2016}
{Binks}, A. 2016, \href{http://dx.doi.org/10.1017/S1743921315006158}{in IAU
  Symposium, Vol. 314, IAU Symposium, ed. J.~H. {Kastner}, B.~{Stelzer}, \&
  S.~A. {Metchev}}, 159

\bibitem[{{Boccaletti} {et~al.}(2015){Boccaletti}, {Thalmann}, {Lagrange},
  {Janson}, {Augereau}, {Schneider}, {Milli}, {Grady}, {Debes}, {Langlois},
  {Mouillet}, {Henning}, {Dominik}, {Maire}, {Beuzit}, {Carson}, {Dohlen},
  {Engler}, {Feldt}, {Fusco}, {Ginski}, {Girard}, {Hines}, {Kasper}, {Mawet},
  {M{\'e}nard}, {Meyer}, {Moutou}, {Olofsson}, {Rodigas}, {Sauvage},
  {Schlieder}, {Schmid}, {Turatto}, {Udry}, {Vakili}, {Vigan}, {Wahhaj}, \&
  {Wisniewski}}]{Boccaletti2015}
{Boccaletti}, A., {Thalmann}, C., {Lagrange}, A.-M., {et~al.} 2015,
  \href{http://dx.doi.org/10.1038/nature15705}{\JournalTitle{\nat}, 526, 230}

\bibitem[{{Brandt} {et~al.}(2014){Brandt}, {Kuzuhara}, {McElwain}, {Schlieder},
  {Wisniewski}, {Turner}, {Carson}, {Matsuo}, {Biller}, {Bonnefoy}, {Dressing},
  {Janson}, {Knapp}, {Moro-Mart{\'{\i}}n}, {Thalmann}, {Kudo}, {Kusakabe},
  {Hashimoto}, {Abe}, {Brandner}, {Currie}, {Egner}, {Feldt}, {Golota}, {Goto},
  {Grady}, {Guyon}, {Hayano}, {Hayashi}, {Hayashi}, {Henning}, {Hodapp},
  {Ishii}, {Iye}, {Kandori}, {Kwon}, {Mede}, {Miyama}, {Morino}, {Nishimura},
  {Pyo}, {Serabyn}, {Suenaga}, {Suto}, {Suzuki}, {Takami}, {Takahashi},
  {Takato}, {Terada}, {Tomono}, {Watanabe}, {Yamada}, {Takami}, {Usuda}, \&
  {Tamura}}]{Brandt2014}
{Brandt}, T.~D., {Kuzuhara}, M., {McElwain}, M.~W., {et~al.} 2014,
  \href{http://dx.doi.org/10.1088/0004-637X/786/1/1}{\JournalTitle{\apj}, 786,
  1}

\bibitem[{{Currie} {et~al.}(2014){Currie}, {Daemgen}, {Debes}, {Lafreniere},
  {Itoh}, {Jayawardhana}, {Ratzka}, \& {Correia}}]{Currie2014}
{Currie}, T., {Daemgen}, S., {Debes}, J., {et~al.} 2014,
  \href{http://dx.doi.org/10.1088/2041-8205/780/2/L30}{\JournalTitle{ApJL},
  780, L30}

\bibitem[{{Currie} {et~al.}(2015){Currie}, {Lisse}, {Kuchner}, {Madhusudhan},
  {Kenyon}, {Thalmann}, {Carson}, \& {Debes}}]{Currie2015}
{Currie}, T., {Lisse}, C.~M., {Kuchner}, M., {et~al.} 2015,
  \href{http://dx.doi.org/10.1088/2041-8205/807/1/L7}{\JournalTitle{ApJL}, 807,
  L7}

\bibitem[{{Cutri} \& {et al.}(2014)}]{Cutri2014}
{Cutri}, R.~M., \& {et al.} 2014, \JournalTitle{VizieR Online Data Catalog},
  2328

\bibitem[{{Epchtein} {et~al.}(1997){Epchtein}, {de Batz}, {Capoani},
  {Chevallier}, {Copet}, {Fouqu{\'e}}, {Lacombe}, {Le Bertre}, {Pau}, {Rouan},
  {Ruphy}, {Simon}, {Tiph{\`e}ne}, {Burton}, {Bertin}, {Deul}, {Habing},
  {Borsenberger}, {Dennefeld}, {Guglielmo}, {Loup}, {Mamon}, {Ng}, {Omont},
  {Provost}, {Renault}, {Tanguy}, {Kimeswenger}, {Kienel}, {Garzon}, {Persi},
  {Ferrari-Toniolo}, {Robin}, {Paturel}, {Vauglin}, {Forveille}, {Delfosse},
  {Hron}, {Schultheis}, {Appenzeller}, {Wagner}, {Balazs}, {Holl},
  {L{\'e}pine}, {Boscolo}, {Picazzio}, {Duc}, \& {Mennessier}}]{Epchtein1997}
{Epchtein}, N., {de Batz}, B., {Capoani}, L., {et~al.} 1997, \JournalTitle{The
  Messenger}, 87, 27

\bibitem[{{ESA}(1997)}]{Hipparcos}
{ESA}, ed. 1997, ESA Special Publication, Vol. 1200, {The HIPPARCOS and TYCHO
  catalogues. Astrometric and photometric star catalogues derived from the ESA
  HIPPARCOS Space Astrometry Mission}

\bibitem[{{Espaillat} {et~al.}(2010){Espaillat}, {D'Alessio}, {Hern{\'a}ndez},
  {Nagel}, {Luhman}, {Watson}, {Calvet}, {Muzerolle}, \&
  {McClure}}]{Espaillat2010}
{Espaillat}, C., {D'Alessio}, P., {Hern{\'a}ndez}, J., {et~al.} 2010,
  \href{http://dx.doi.org/10.1088/0004-637X/717/1/441}{\JournalTitle{\apj},
  717, 441}

\bibitem[{{Gagn{\'e}} {et~al.}(2014){Gagn{\'e}}, {Lafreni{\`e}re}, {Doyon},
  {Malo}, \& {Artigau}}]{Gagne2014}
{Gagn{\'e}}, J., {Lafreni{\`e}re}, D., {Doyon}, R., {Malo}, L., \& {Artigau},
  {\'E}. 2014,
  \href{http://dx.doi.org/10.1088/0004-637X/783/2/121}{\JournalTitle{\apj},
  783, 121}

\bibitem[{{Girard} {et~al.}(2011){Girard}, {van Altena}, {Zacharias}, {Vieira},
  {Casetti-Dinescu}, {Castillo}, {Herrera}, {Lee}, {Beers}, {Monet}, \&
  {L{\'o}pez}}]{Girard2011}
{Girard}, T.~M., {van Altena}, W.~F., {Zacharias}, N., {et~al.} 2011,
  \href{http://dx.doi.org/10.1088/0004-6256/142/1/15}{\JournalTitle{\aj}, 142,
  15}

\bibitem[{{Heng} \& {Malik}(2013)}]{HengMalik2013}
{Heng}, K., \& {Malik}, M. 2013,
  \href{http://dx.doi.org/10.1093/mnras/stt615}{\JournalTitle{\mnras}, 432,
  2562}

\bibitem[{{H{\o}g} {et~al.}(2000){H{\o}g}, {Fabricius}, {Makarov}, {Urban},
  {Corbin}, {Wycoff}, {Bastian}, {Schwekendiek}, \& {Wicenec}}]{Hog2000}
{H{\o}g}, E., {Fabricius}, C., {Makarov}, V.~V., {et~al.} 2000,
  \JournalTitle{\aap}, 355, L27

\bibitem[{{Houk}(1982)}]{Houk1982}
{Houk}, N. 1982, {Michigan Catalogue of Two-dimensional Spectral Types for the
  HD stars. Volume 3. Declinations -40\_0 to -26\_0.}

\bibitem[{{Kasper} {et~al.}(2007){Kasper}, {Apai}, {Janson}, \&
  {Brandner}}]{Kasper2007}
{Kasper}, M., {Apai}, D., {Janson}, M., \& {Brandner}, W. 2007,
  \href{http://dx.doi.org/10.1051/0004-6361:20077646}{\JournalTitle{\aap}, 472,
  321}

\bibitem[{{Kennedy} \& {Wyatt}(2010)}]{KennedyWyatt2010}
{Kennedy}, G.~M., \& {Wyatt}, M.~C. 2010,
  \href{http://dx.doi.org/10.1111/j.1365-2966.2010.16528.x}{\JournalTitle{\mnras},
  405, 1253}

\bibitem[{{Kennedy} \& {Wyatt}(2013)}]{KennedyWyatt2013}
---. 2013, \href{http://dx.doi.org/10.1093/mnras/stt900}{\JournalTitle{\mnras},
  433, 2334}

\bibitem[{{Kuchner} {et~al.}(2016){Kuchner}, {Silverberg}, {Bans},
  {Bhattacharjee}, {Kenyon}, {Debes}, {Currie}, {Garcia}, {Jung}, {Lintott},
  {McElwain}, {Padgett}, {Rebull}, {Wisniewski}, {Nesvold}, {Schawinski},
  {Thaller}, {Grady}, {Biggs}, {Bosch}, {Cernohous}, {Durantini-Luca}, {Hyogo},
  {Lau~Wan~Wah}, {Piipuu}, \& {Piniero}}]{Kuchner2016}
{Kuchner}, M.~J., {Silverberg}, S.~M., {Bans}, A.~S., {et~al.} 2016,
  \JournalTitle{\aj}

\bibitem[{{Lindegren, L.} {et~al.}(2016){Lindegren, L.}, {Lammers, U.},
  {Bastian, U.}, {HernÃ¡ndez, J.}, {Klioner, S.}, {Hobbs, D.}, {Bombrun, A.},
  \& {Michalik, D.}}]{Lindegren2016}
{Lindegren, L.}, {Lammers, U.}, {Bastian, U.}, {et~al.} 2016,
  \href{http://dx.doi.org/10.1051/0004-6361/201628714}{\JournalTitle{\aap}}

\bibitem[{{L{\'o}pez-Santiago} {et~al.}(2006){L{\'o}pez-Santiago}, {Montes},
  {Crespo-Chac{\'o}n}, \& {Fern{\'a}ndez-Figueroa}}]{LopezSantiago2006}
{L{\'o}pez-Santiago}, J., {Montes}, D., {Crespo-Chac{\'o}n}, I., \&
  {Fern{\'a}ndez-Figueroa}, M.~J. 2006,
  \href{http://dx.doi.org/10.1086/503183}{\JournalTitle{\apj}, 643, 1160}

\bibitem[Luhman \& Mamajek(2012)]{LuhmanMamajek2012} Luhman, K.~L., \& Mamajek, E.~E.\ 2012, \apj, 758, 31 

\bibitem[{{Malo} {et~al.}(2013){Malo}, {Doyon}, {Lafreni{\`e}re}, {Artigau},
  {Gagn{\'e}}, {Baron}, \& {Riedel}}]{Malo2013}
{Malo}, L., {Doyon}, R., {Lafreni{\`e}re}, D., {et~al.} 2013,
  \href{http://dx.doi.org/10.1088/0004-637X/762/2/88}{\JournalTitle{\apj}, 762,
  88}

\bibitem[{{Marois} {et~al.}(2008){Marois}, {Macintosh}, {Barman}, {Zuckerman},
  {Song}, {Patience}, {Lafreni{\`e}re}, \& {Doyon}}]{Marois2008}
{Marois}, C., {Macintosh}, B., {Barman}, T., {et~al.} 2008,
  \href{http://dx.doi.org/10.1126/science.1166585}{\JournalTitle{Science}, 322,
  1348}

\bibitem[{{Marois} {et~al.}(2010){Marois}, {Zuckerman}, {Konopacky},
  {Macintosh}, \& {Barman}}]{Marois2010}
{Marois}, C., {Zuckerman}, B., {Konopacky}, Q.~M., {Macintosh}, B., \&
  {Barman}, T. 2010,
  \href{http://dx.doi.org/10.1038/nature09684}{\JournalTitle{\nat}, 468, 1080}

\bibitem[{{Perryman} {et~al.}(2001){Perryman}, {de Boer}, {Gilmore}, {H{\o}g},
  {Lattanzi}, {Lindegren}, {Luri}, {Mignard}, {Pace}, \& {de
  Zeeuw}}]{Perryman2001}
{Perryman}, M.~A.~C., {de Boer}, K.~S., {Gilmore}, G., {et~al.} 2001,
  \href{http://dx.doi.org/10.1051/0004-6361:20010085}{\JournalTitle{\aap}, 369,
  339}

\bibitem[{{Plavchan} {et~al.}(2005){Plavchan}, {Jura}, \&
  {Lipscy}}]{Plavchan2005}
{Plavchan}, P., {Jura}, M., \& {Lipscy}, S.~J. 2005,
  \href{http://dx.doi.org/10.1086/432568}{\JournalTitle{\apj}, 631, 1161}

\bibitem[{{Plavchan} {et~al.}(2009){Plavchan}, {Werner}, {Chen}, {Stapelfeldt},
  {Su}, {Stauffer}, \& {Song}}]{Plavchan2009}
{Plavchan}, P., {Werner}, M.~W., {Chen}, C.~H., {et~al.} 2009,
  \href{http://dx.doi.org/10.1088/0004-637X/698/2/1068}{\JournalTitle{\apj},
  698, 1068}

\bibitem[{{Rajpurohit} {et~al.}(2013){Rajpurohit}, {Reyl{\'e}}, {Allard},
  {Homeier}, {Schultheis}, {Bessell}, \& {Robin}}]{Rajpurohit2013}
{Rajpurohit}, A.~S., {Reyl{\'e}}, C., {Allard}, F., {et~al.} 2013,
  \href{http://dx.doi.org/10.1051/0004-6361/201321346}{\JournalTitle{\aap},
  556, A15}

\bibitem[{{Rizzuto} {et~al.}(2012){Rizzuto}, {Ireland}, \&
  {Zucker}}]{Rizzuto2012}
{Rizzuto}, A.~C., {Ireland}, M.~J., \& {Zucker}, D.~B. 2012,
  \href{http://dx.doi.org/10.1111/j.1745-3933.2012.01214.x}{\JournalTitle{\mnras},
  421, L97}

\bibitem[{{Schneider} {et~al.}(2012{\natexlab{a}}){Schneider}, {Melis}, \&
  {Song}}]{Schneider2012a}
{Schneider}, A., {Melis}, C., \& {Song}, I. 2012{\natexlab{a}},
  \href{http://dx.doi.org/10.1088/0004-637X/754/1/39}{\JournalTitle{\apj}, 754,
  39}

\bibitem[{{Schneider} {et~al.}(2012{\natexlab{b}}){Schneider}, {Song}, {Melis},
  {Zuckerman}, \& {Bessell}}]{Schneider2012b}
{Schneider}, A., {Song}, I., {Melis}, C., {Zuckerman}, B., \& {Bessell}, M.
  2012{\natexlab{b}},
  \href{http://dx.doi.org/10.1088/0004-637X/757/2/163}{\JournalTitle{\apj},
  757, 163}

\bibitem[{{Schneider} {et~al.}(2014){Schneider}, {Grady}, {Hines}, {Stark},
  {Debes}, {Carson}, {Kuchner}, {Perrin}, {Weinberger}, {Wisniewski},
  {Silverstone}, {Jang-Condell}, {Henning}, {Woodgate}, {Serabyn},
  {Moro-Martin}, {Tamura}, {Hinz}, \& {Rodigas}}]{Schneider2014}
{Schneider}, G., {Grady}, C.~A., {Hines}, D.~C., {et~al.} 2014,
  \href{http://dx.doi.org/10.1088/0004-6256/148/4/59}{\JournalTitle{\aj}, 148,
  59}

\bibitem[{{Skrutskie} {et~al.}(2006){Skrutskie}, {Cutri}, {Stiening},
  {Weinberg}, {Schneider}, {Carpenter}, {Beichman}, {Capps}, {Chester},
  {Elias}, {Huchra}, {Liebert}, {Lonsdale}, {Monet}, {Price}, {Seitzer},
  {Jarrett}, {Kirkpatrick}, {Gizis}, {Howard}, {Evans}, {Fowler}, {Fullmer},
  {Hurt}, {Light}, {Kopan}, {Marsh}, {McCallon}, {Tam}, {Van Dyk}, \&
  {Wheelock}}]{Skrutskie2006}
{Skrutskie}, M.~F., {Cutri}, R.~M., {Stiening}, R., {et~al.} 2006,
  \href{http://dx.doi.org/10.1086/498708}{\JournalTitle{\aj}, 131, 1163}

\bibitem[{{Su} {et~al.}(2006){Su}, {Rieke}, {Stansberry}, {Bryden},
  {Stapelfeldt}, {Trilling}, {Muzerolle}, {Beichman}, {Moro-Martin}, {Hines},
  \& {Werner}}]{Su2006}
{Su}, K.~Y.~L., {Rieke}, G.~H., {Stansberry}, J.~A., {et~al.} 2006,
  \href{http://dx.doi.org/10.1086/508649}{\JournalTitle{\apj}, 653, 675}

\bibitem[{{Theissen} \& {West}(2014)}]{TheissenWest2014}
{Theissen}, C.~A., \& {West}, A.~A. 2014,
  \href{http://dx.doi.org/10.1088/0004-637X/794/2/146}{\JournalTitle{\apj},
  794, 146}

\bibitem[{{Torres} {et~al.}(2006){Torres}, {Quast}, {da Silva}, {de La Reza},
  {Melo}, \& {Sterzik}}]{Torres2006}
{Torres}, C.~A.~O., {Quast}, G.~R., {da Silva}, L., {et~al.} 2006,
  \href{http://dx.doi.org/10.1051/0004-6361:20065602}{\JournalTitle{\aap}, 460,
  695}

\bibitem[{{Torres} {et~al.}(2003){Torres}, {Quast}, {de La Reza}, {da Silva},
  {Melo}, \& {Sterzik}}]{Torres2003}
{Torres}, C.~A.~O., {Quast}, G.~R., {de La Reza}, R., {et~al.} 2003,
  \href{http://dx.doi.org/10.1007/978-1-4020-2600-3_12}{in Astrophysics and
  Space Science Library, Vol. 299, Astrophysics and Space Science Library, ed.
  J.~{L{\'e}pine} \& J.~{Gregorio-Hetem}}, 83

\bibitem[{{Torres} {et~al.}(2008){Torres}, {Quast}, {Melo}, \&
  {Sterzik}}]{Torres2008}
{Torres}, C.~A.~O., {Quast}, G.~R., {Melo}, C.~H.~F., \& {Sterzik}, M.~F. 2008,
  {Young Nearby Loose Associations}, ed. B.~{Reipurth}, 757

\bibitem[{{Urban} {et~al.}(2012){Urban}, {Rieke}, {Su}, \&
  {Trilling}}]{Urban2012}
{Urban}, L.~E., {Rieke}, G., {Su}, K., \& {Trilling}, D.~E. 2012,
  \href{http://dx.doi.org/10.1088/0004-637X/750/2/98}{\JournalTitle{\apj}, 750,
  98}

\bibitem[Williams \& Cieza(2011)]{WilliamsCieza2011} Williams, J.~P., \& Cieza, L.~A.\ 2011, \araa, 49, 67 

\bibitem[{{Wright} {et~al.}(2010){Wright}, {Eisenhardt}, {Mainzer}, {Ressler},
  {Cutri}, {Jarrett}, {Kirkpatrick}, {Padgett}, {McMillan}, {Skrutskie},
  {Stanford}, {Cohen}, {Walker}, {Mather}, {Leisawitz}, {Gautier}, {McLean},
  {Benford}, {Lonsdale}, {Blain}, {Mendez}, {Irace}, {Duval}, {Liu}, {Royer},
  {Heinrichsen}, {Howard}, {Shannon}, {Kendall}, {Walsh}, {Larsen}, {Cardon},
  {Schick}, {Schwalm}, {Abid}, {Fabinsky}, {Naes}, \& {Tsai}}]{Wright2010}
{Wright}, E.~L., {Eisenhardt}, P.~R.~M., {Mainzer}, A.~K., {et~al.} 2010,
  \href{http://dx.doi.org/10.1088/0004-6256/140/6/1868}{\JournalTitle{\aj},
  140, 1868}

\bibitem[{{Zuckerman} {et~al.}(2011){Zuckerman}, {Rhee}, {Song}, \&
  {Bessell}}]{Zuckerman2011}
{Zuckerman}, B., {Rhee}, J.~H., {Song}, I., \& {Bessell}, M.~S. 2011,
  \href{http://dx.doi.org/10.1088/0004-637X/732/2/61}{\JournalTitle{\apj}, 732,
  61}

\bibitem[{{Zuckerman} \& {Song}(2004)}]{ZuckermanSong2004}
{Zuckerman}, B., \& {Song}, I. 2004,
  \href{http://dx.doi.org/10.1146/annurev.astro.42.053102.134111}{\JournalTitle{\araa},
  42, 685}

\end{thebibliography}
\end{document}